\documentclass{article}
\usepackage{graphics}
\def\abstract#1{\vskip 7mm
        \begin{center}{\large Abstract}\par \smallskip
                \begin{minipage}[c]{12cm}
                        \small #1
                \end{minipage}
        \end{center}
}
\def\title#1{\begin{center}{\Large\bf #1}\end{center}}
\def\author#1{\vskip 5mm \begin{center}{#1}\end{center}}
\def\address#1{\begin{center}{\it #1}\end{center}}
\makeatletter
\@ifundefined{lesssim}{}{}
\@ifundefined{gtrsim}{}{}
\def\vereq#1#2{\lower3pt\vbox{\baselineskip1.5pt \lineskip1.5pt
\ialign{$\m@th#1\hfill##\hfil$\crcr#2\crcr\sim\crcr}}}
\makeatother

\begin{document}
\begin{flushright}
Sofia University\\
\end{flushright}
%------------------------------------------------------------------------------
%
%------------------------------------------------------------------------------
\title{Non-asymptotically flat, non-dS/AdS dyonic  black holes in dilaton gravity }
\author{%
  Stoytcho S. Yazadjiev\footnote{E-mail:yazad@phys.uni-sofia.bg}
}
\address{%
  Department of Theoretical Physics,
  Faculty of Physics, Sofia University,\\
  5 James Bourchier Boulevard, Sofia~1164, Bulgaria
}
%------------------------------------------------------------------------------
%
%------------------------------------------------------------------------------
\abstract{We present exact spherically symmetric  dyonic black hole solutions in four-dimensional
and higher dimensional Einstein-Maxwell-dilaton gravity with Liouville-type potentials for the dilaton field.
These solutions have unusual asymptotics--they are neither asymptotically flat nor asymptotically
(anti-) de Sitter. The solutions have one or two horizons hiding a curvature singularity at the origin.
A class of topological dyonic black holes with topology of a torus is also presented in four dimensions.
Some basic properties of the black holes are discussed.}
%------------------------------------------------------------------------------
%
%------------------------------------------------------------------------------
\section{Introduction and general equations}

In recent years non-asymptotically flat black hole spacetimes are attracting much interest
in connection with the so called AdS/CFT correspondence. Moreover, black hole spacetimes which are neither
asymptotically flat nor dS/AdS have been found and investigated, too. The first such uncharged solutions were found by Mignemi and Wiltshire in \cite{MW1}, Wiltshire in \cite{W} and Mignemi and Wiltshire in \cite{MW2}. Charged
black hole solutions in Einstein-Maxwell-dilaton (EMd) gravity with a Liouville potential were first obtained
by Poletti and Wiltshire in \cite{PW}. Chan, Horne and Mann \cite{CHM} presented exact static non-asymptotically flat, non-dS/AdS black holes with electric (or magnetic) charge in $n$-dimensional EMd  gravity with certain  Liouville-type potentials. In four dimensions, when the two-sphere of the black hole horizon is replaced by a two-dimensional hypersurface  with zero or negative constant curvature (the so-called topological black holes), the
solutions of \cite{PW} and \cite{CHM} were generalized by Cai, Ji and Soh \cite{CJS} (see also \cite{CZ}). Using the $Sp(4,R)$ duality of the four dimensional Einstein-Maxwell-dilaton-axion gravity  Clement, Gal'tsov and Leygnac generated exact rotating dilaton-axion black holes in linear dilaton background \cite{CGL}.
Rotating non-AdS dyonic black hole  solutions with a specialized dilaton coupling parameter $\beta^2=3$ were constructed by Clement and Leygnac in \cite{CL}. Static, electrically or magnetically charged, non-asymptotically flat, non-dS/AdS black holes in various dimensions were also found and discussed  in the work by Cai and Wang \cite{CW}.

In this  paper we obtain and discuss exact black hole solutions in four-dimensional and higher dimensional EMd gravity. These solutions posses both electric and magnetic charge and they are neither asymptotically flat nor asymptotically dS/AdS. The asymptotically flat dyonic black holes were considered in \cite{GM}, \cite{KLOPVP},\cite{ALC} and \cite{CLH}. The paper is organized as follows. First we present and discuss EMd black holes in four dimensions. Section VI is devoted to the black hole solutions in $n$-dimensional spacetimes with $n>4$.

The four dimensional EMd gravity with a dilaton potential is described by the action

\begin{equation}\label{EMDA}
{\cal S} = \int d^4x \sqrt{-g}\left[{\cal R} - 2g^{\mu\nu}\partial_{\mu}\varphi \partial_{\nu}\varphi - e^{-2\beta\varphi}F_{\mu\nu}F^{\mu\nu}  - V(\varphi)\right]
\end{equation}

where $g_{\mu\nu}$ is the spacetime metric, $F$ is the Maxwell field, and $V(\varphi)$ is a  potential for the dilaton field  $\varphi$.  The dilaton coupling parameter is denoted by $\beta$ ($\beta \ne 0$). The action (\ref{EMDA}) yields
the following field equations

\begin{eqnarray}
{\cal R}_{\mu\nu} = 2\partial_{\mu}\varphi \partial_{\nu}\varphi + 2e^{-2\beta\varphi}\left(F_{\mu\alpha}F_{\nu}^{\alpha}-
{1\over 4} g_{\mu\nu} F^2 \right) + {1\over 2}V(\varphi)g_{\mu\nu},\\
\nabla_{\mu}\left(e^{-2\beta\varphi }F^{\mu\nu} \right)= 0 ,\\
\nabla_{\mu}\nabla^{\mu}\varphi = -{\beta\over 2}e^{-2\beta\varphi} F^2 + {1\over 4} {dV(\varphi)\over d\varphi }.
\end{eqnarray}

We consider static spherically symmetric  spacetimes with a metric

\begin{equation}
ds^2 = -\lambda(r)dt^2 + {dr^2 \over \lambda(r)} + R^2(r)\left(d\theta^2 + \sin^2\theta d\phi^2 \right).
\end{equation}

The Maxwell equations give

\begin{eqnarray}
F_{rt} = {Qe^{2\beta\varphi} \over R^2(r)}, \\
F_{\theta\phi} = P\sin\theta .
\end{eqnarray}

The electric and magnetic charges are denoted by $Q$ and $P$ and they are defined by the integrals

\begin{eqnarray}
Q &=& {1\over 4\pi} \oint_{S^2} e^{-2\beta\varphi}\star F , \\
P &=& {1\over 4\pi} \oint_{S^2} e^{2\beta\varphi} F ,
\end{eqnarray}

where $\star$ is the Hodge dual and $S^{2}$ is any two-dimensional sphere.

The equations of motion reduce to the following system of ordinary differential equations

\begin{eqnarray}
{1\over R} {d^2R\over  dr^2 }  &=&
- \left({d\varphi\over dr} \right)^2 , \\
{1\over R^2} {d\over dr } \left( \lambda {dR^2\over dr}\right) &=& {2\over R^2} - {2\over R^4}\left(Q^2 e^{2\beta\varphi} + P^2 e^{-2\beta\varphi} \right) - V(\varphi) ,\\
{1\over R^2} {d\over dr }\left(R^2\lambda {d\varphi\over dr } \right) &=& {1\over 4} {dV(\varphi)\over d\varphi } +
{\beta\over R^4}\left(Q^2 e^{2\beta\varphi} - P^2 e^{-2\beta\varphi} \right).
\end{eqnarray}

We make the ansatz

\begin{equation}\label{ansatz}
R(r) = \gamma r^{N}
\end{equation}

where $\gamma$ and $N$ are constants. By making this ansatz many exact solutions were found in the pure electric or
magnetic case \cite{CHM},\cite{CJS},\cite{CW}.

\section{Black holes with a string coupling}

\subsection{Black holes with $V(\varphi)=0$}

We first consider the string coupling case $\beta=1$ with $V(\varphi)=0$. In this case
we have found the following solution

\begin{eqnarray}
N &=& {1\over 2}, \\
\varphi &=& {1\over 2 }\ln(r) - {1\over 2}\ln\left({2Q^2\over \gamma^2 } \right), \\
\lambda(r) &=& {2A\over \gamma^2} + {r\over \gamma^2} + {4Q^2P^2\over \gamma^6 r},
\end{eqnarray}

where $A$ is a constant. Another solution with the same metric can be obtained via the discrete transformation
$Q\longleftrightarrow P$ and $\varphi \longleftrightarrow -\varphi$.

This metric and the other metrics that we will present in this paper are not
asymptotically flat and therefore we must use the so called quasilocal formalism \cite{BY} to define the
mass of the solutions.The quasilocal mass is given by

\begin{equation}\label{QLM}
{\cal M}(r) = {1\over 2} {dR^2(r)\over dr} \lambda^{1/2}(r)\left[\lambda_{0}^{1/2}(r) -\lambda^{1/2}(r) \right]
\end{equation}

where $\lambda_{0}(r)$ is an arbitrary non-negative function which determines the zero of the energy for a background spacetime. If no cosmological horizon is present, the large $r$ limit of (\ref{QLM}) determines the mass.

For the solution under consideration, there is no cosmological horizon and the natural choice for the background
is $\lambda_{0}(r)= r/\gamma^2$.  First we consider the case with $A<0$.  The large $r$ limit of (\ref{QLM})
gives the mass of the solution

\begin{equation}
M = - {A\over 2}
\end{equation}

i.e. the metric function $\lambda(r)$ becomes

\begin{equation}
\lambda(r) = - {4M\over \gamma^2} + {r\over \gamma^2} + {4Q^2P^2\over \gamma^6 r} .
\end{equation}

For $M^2 > Q^2P^2/\gamma^4$, there are two zeros of $\lambda(r)$ at $r=r_{\pm}$:

\begin{equation}
r_{\pm} = 2\left( M \pm \sqrt{M^2 - {Q^2P^2\over \gamma^4 }} \right)
\end{equation}

which correspond to two horizons. The scalar invariants are finite at $r_{\pm}$ and diverge at $r=0$,
indicating that $r=r_{\pm}$ are regular horizons and the singularity is located at $r=0$.
As an illustration we present the Ricci scalar curvature

\begin{equation}
{\cal R} = {1\over 2\gamma^2 r } \left(1 - {4M\over r} + {4Q^2P^2 \over \gamma^4 r^2} \right).
\end{equation}

The spacetime has the same causal
structure as that of the Reissner-Nordstrom solution for $M>|Q|$. The regular event horizon is at $r=r_{+}$ while $r=r_{-}$
corresponds to the inner horizon. The temperature of the black hole is given by

\begin{equation}
T = {1\over 4\pi}{d\lambda\over dr}(r_{+}) = {\sqrt{M^2 - {Q^2P^2\over \gamma^4 }} \over 2\pi\gamma^2 \left[ M + \sqrt{M^2 - {Q^2P^2\over \gamma^4 }} \right] }
\end{equation}

while the entropy $S$ is a quarter of the horizon area

\begin{equation}
S = \pi\gamma^2r_{+} = 2\pi\gamma^2 \left( M + \sqrt{M^2 - {Q^2P^2\over \gamma^4 }} \right) .
\end{equation}

The extremal solution corresponds to $M^2= Q^2P^2/\gamma^4$. In this case $\lambda(r)$ has only one root at
$r=r_{ext}$:

\begin{equation}
r_{ext} = 2M = 2{|QP|\over \gamma^2}.
\end{equation}

The horizon $r_{ext}$ is a regular event horizon formed by merging of the horizons $r=r_{-}$ and $r=r_{+}$
while $r=0$ is still a singularity.

The solutions with $M^2 < Q^2P^2/\gamma^4$ are naked singularities. The same holds for $A\ge0$.

\subsection{Black holes with a Liouville  potential}

Here we consider black holes in EMd gravity with $\beta=1$ and  Liouville  potential

\begin{equation}
V=\Lambda e^{2k\varphi}
\end{equation}

where $\Lambda$ and $k$ are constants.

We have found the following solution:

\begin{eqnarray}
N &=& {1\over 2},\\
k &=& -1, \\
\varphi(r) &=& {1\over 2} \ln(r) - {1\over 2}\ln\left( {2Q^2\over \gamma^2}\right),\\
\lambda(r) &=& {2A\over \gamma^2} + {1 -2\Lambda Q^2\over \gamma^2 } r + {4Q^2P^2\over \gamma^6 r } .
\end{eqnarray}

Another solution with the same spacetime metric can be obtained via the discrete transformation
$Q \longleftrightarrow P$, $ \varphi \longleftrightarrow - \varphi$, $k\longleftrightarrow -k$.

The scalar curvature of the metric is

\begin{equation}\label{SCLSC}
{\cal R} = {1\over 2\gamma^2 r } \left(1 + 6\Lambda Q^2- {4M\over r} + {4Q^2P^2 \over \gamma^4 r^2} \right).
\end{equation}

The three cases $2\Lambda Q^2 <1$, $2\Lambda Q^2 >1$ and $2\Lambda Q^2=1$ should be considered separately.

\subsection*{Solutions with $2\Lambda Q^2 <1$}

In this case there is no cosmological horizon and eq.(\ref{QLM}) gives for the the mass $M= -{A\over 2}$
where we have taken $\lambda_{0}(r) = [(1 -2\Lambda Q^2)/\gamma^2] r$ and $A<0$. Therefore we have

\begin{equation}
 \lambda(r) = - {4M\over \gamma^2} + {1 -2\Lambda Q^2\over \gamma^2 } r + {4Q^2P^2\over \gamma^6 r }.
\end{equation}

For $M^2 > {Q^2P^2\over \gamma^4}(1-2\Lambda Q^2)$, the function $\lambda(r)$ has two zeros

\begin{equation}
r_{\pm} = {2\over 1- 2\Lambda Q^2} \left[M \pm \sqrt{M^2 - {Q^2P^2\over\gamma^4 }(1-2\Lambda Q^2)} \right]
\end{equation}

where $r_{+}$ corresponds to a regular event horizon and $r=r_{-}$ is the location of the inner regular horizon.

The curvature invariants diverge at $r=0$ as one can see from (\ref{SCLSC}) in the particular case of the
scalar curvature.

The black hole temperature is

\begin{equation}
T = {1\over 2\pi\gamma^2}(1-2\Lambda Q^2){\sqrt{M^2 - {Q^2P^2\over\gamma^4 }(1-2\Lambda Q^2)}\over \left[M + \sqrt{M^2 - {Q^2P^2\over\gamma^4 }(1-2\Lambda Q^2)} \right] }
\end{equation}

while the entropy is given by

\begin{equation}
S = \pi \gamma^2 r_{+} = {2\pi\gamma^2\over 1- 2\Lambda Q^2} \left[M \pm \sqrt{M^2 - {Q^2P^2\over\gamma^4 }(1-2\Lambda Q^2)} \right] .
\end{equation}

For $M^2= {Q^2P^2\over \gamma^4}(1-2\Lambda Q^2)$, $\lambda(r)$ has only one zero
\begin{equation}
r_{ext}= {2M \over (1-2\Lambda Q^2)} = {2|QP|\over \gamma^2 \sqrt{1 - 2\Lambda Q^2}}
\end{equation}
corresponding to a  degenerate horizon hiding a singularity at $r=0$.

The solutions with $M^2 < {Q^2P^2\over \gamma^4}(1-2\Lambda Q^2)$  and $A\ge0$ are naked singularities.

\subsection*{Solutions with $2\Lambda Q^2 >1$}

In this case there is a cosmological horizon at

\begin{equation}
r_{CH} = {A + \sqrt{A^2 + {4Q^2P^2\over \gamma^4 } |1-2\Lambda Q^2|}\over |1 -2\Lambda Q^2| }.
\end{equation}

\subsection*{Solutions with $2\Lambda Q^2 =1$}

For $A<0$ a cosmological horizon is present at

\begin{equation}
r_{CH} = {2Q^2P^2\over \gamma^4 |A| }.
\end{equation}

The case  $A\ge 0$ corresponds to a massless naked singularity at $r=0$.

\section{Black holes with a general coupling parameter and Liouville potential  }

In this section we present exact solutions of EMd gravity with an arbitrary dilaton
coupling parameter $\beta$ and a Liouville potential $V(\varphi)= \Lambda e^{2k\varphi}$.

The solution is given by:

\begin{eqnarray}
N &=&{\beta^2 \over 1 + \beta^2},\\
k &=& -3\beta,\\
\varphi(r) &=& {\beta\over 1 + \beta^2} \ln(r) - {1\over 2\beta}\ln\left[{(1+\beta^2)Q^2 \over \gamma^2} \right],\\
\lambda(r) &=& {r^{1-2N}\over \gamma^2}\left[ {A(1+\beta^2)\over \beta } +r + {2(1+\beta^2)^{3}\over (1-3\beta^2)^2 } {Q^2P^2\over \gamma^4 } r^{1-4N} \right],\\
\Lambda &=& - {2(1-\beta^2)\over (1-3\beta^2) (1+\beta^2)^2} {P^2\over Q^4 } .
 \end{eqnarray}

Another solution with the same spacetime metric is generated via the discrete transformation
$Q \longleftrightarrow P$, $\varphi \longleftrightarrow -\varphi$, $k \longleftrightarrow -k$.

The solution is ill defined for $\beta^2={1/3}$. In the particular case $\beta^2=1$ the solution reduces to that
with $V(\varphi)=0$. There is no cosmological horizon and  the mass can be computed from (\ref{QLM}). For the
background function $\lambda_{0}(r) = r^{2-2N}/\gamma^2$ and $\beta A<0$ the mass is:

\begin{equation}
M = -{\beta\over 2} A
\end{equation}

and therefore

\begin{equation}
\lambda(r) = {r^{1-2N}\over \gamma^2}\left[r - {2(1+\beta^2)M\over \beta^2} + {2(1+\beta^2)^{3}\over (1-3\beta^2)^2 } {Q^2P^2\over \gamma^4 } r^{1-4N} \right].
\end{equation}

The curvature invariants are regular except for $r=0$ where they diverge. This can be seen from the explicit expression of the scalar curvature ${\cal R}$:

\begin{equation}
{\cal R} = {2\beta^2\over (1+\beta^2)^2 } {r^{-2N}\over \gamma^2 } \left[1 + {A(1+ \beta^2) \over \gamma^2 r } -
{2(1+\beta^2)^3\over (1-3\beta^2)^2 \beta^2 }(1-5\beta^2 + 3\beta^4) {Q^2P^2\over \gamma^4 }r^{-4N} \right] .
\end{equation}

In order to investigate the causal structure of the solution we must investigate  the zeros of the metric function
$\lambda(r)$. In fact, for $0<r <\infty$ the zeros of $\lambda(r)$ are governed by the function

\begin{equation}
f(r) = - {2(1+\beta^2)M\over \beta^2 } + r + {2(1+\beta^2)^{3}\over (1-3\beta^2)^2 } {Q^2P^2\over \gamma^4 } r^{1-4N}.
\end{equation}

The cases $\beta^2 <1/3$ and $\beta^2 >1/3$ should be considered separately. In the first case, when $\beta^2 <1/3$
the function $f(r)$ is monotonically increasing (${df\over dr }>0$). In addition $f(r)<0$ for small enough values of $r$ and $f(r)>0$ for large enough values of $r$. Therefore, $\lambda(r)$ has only one zero,i.e. the black hole has one horizon. Unfortunately, the black hole radius determined by $f(r)=0$ can not be expressed in a closed analytical form for arbitrary $\beta$.

In the second case, $\beta^2 >1/3$, the function $f(r)$ satisfies $f(r)\to + \infty$ for
$r \to 0$ and $r\to \infty$. There is only one local minimum at $r=r_{min}$  where

\begin{equation}
r_{min} = \left[{2(1+\beta^2)^2\over (3\beta^2 -1) } {Q^2P^2\over \gamma^4}\right]^{1 +\beta^2 \over 4\beta^2 }.
\end{equation}

The function $f(r)$ possesses zeros only when $f(r_{min}) \le 0$. There are two zeros
for $f(r_{min})<0$ and only one degenerate zero for $f(r_{min})=0$ which corresponds to an extremal black hole. The condition
$f(r_{min}) \le 0$ gives

\begin{equation}\label{MCE}
M \ge {2\beta^4\over (1+\beta^2)(3\beta^2 -1) } \left[ {2(1+\beta^2)^2 \over (3\beta^2-1)}{Q^2P^2\over \gamma^4 }\right]^{1+\beta^2\over 4\beta^2 }
\end{equation}

where the equality holds for the extremal case.

The above considerations show that the solutions describe black holes with two horizons or an extremal black hole hiding a singularity at the origin $r=0$, when the mass and charges satisfy (\ref{MCE}). The radius of the inner and outer horizons can not be expressed in a closed analytical form except for the extremal case. The radius of the extremal solution coincides with $r_{min}$:

\begin{equation}
r_{ext}=r_{min} = \left[{2(1+\beta^2)^2\over (3\beta^2 -1) } {Q^2P^2\over \gamma^4}\right]^{1 +\beta^2 \over 4\beta^2 } = {(1+\beta^2)(3\beta^2-1)\over  2\beta^4} M .
\end{equation}

The solutions which do not satisfy (\ref{MCE}) and those with $\beta A\ge 0$ are  naked singularities.

\section{Solutions with a general coupling parameter and two Liouville potentials}

Here we present exact solution to the EMd gravity equations with an arbitrary dilaton coupling parameter
$\beta$ and dilaton potential $V(\varphi) = \Lambda_{1} e^{2k_{1}\varphi} + \Lambda_{2} e^{2k_{2}\varphi}$.

We have found the solution

\begin{eqnarray}
N &=& {\beta^2\over 1 + \beta^2},\\
k_{1} &=& -\beta ,\\
k_{2} &=& -3\beta ,\\
\Lambda_{1} &=& {2e^{2\beta\varphi_{0}}\over \gamma^2 (1-\beta^2) } \left[1 - (1+\beta^2){Q^2 e^{2\beta\varphi_{0}}\over \gamma^2} \right],\\
\Lambda_{2} &=& -2 {1-\beta^2\over 1-3\beta^2 } {P^2e^{4\beta\varphi_{0}}\over \gamma^4},\\
\varphi(r) &=& {\beta\over 1+\beta^2 }\ln(r) + \varphi_{0},\\
\lambda(r) &=& {r^{1-2N}\over \gamma^2 } \left[ - {A(1+\beta^2)\over \beta }   + {1+\beta^2\over 1-\beta^2 }\left({2Q^2 e^{2\beta\varphi_{0}} \over \gamma^2 } - 1 \right) r  + 2 \left({1+\beta^2\over 1-3\beta^2 } \right)^2 {P^2e^{-2\beta\varphi_{0}}\over \gamma^2}  r^{1-4N}\right],
\end{eqnarray}

where $A$ and $\varphi_{0}$ are arbitrary constants. The discrete transformation $Q \longleftrightarrow  P$,
$\varphi \longleftrightarrow -\varphi$, $k \longleftrightarrow -k $ generates another solution with the same spacetime metric. The solution is not defined for $\beta^2=1$ and $\beta^2=1/3$.

The curvature scalars  are divergent at $r=0$ and regular for $r>0$. As an illustration we present the Ricci scalar
curvature:

\begin{equation}
{\cal R} = {2\beta^2\over (1+\beta^2) } {r^{-2N}\over \gamma^2 } \left[{1+ \beta^2\over (1-\beta^2)\beta^2}
\left( 2 - \beta^2 -2 {Q^2e^{2\beta\varphi_{0}}\over \gamma^2 }\right)  - {A\over \beta r } -
2 {(1 - 5\beta^2 + 2\beta^4)\over (1-3\beta^2)\beta^2 } {P^2e^{-2\beta\varphi_{0}}\over \gamma^2 } r^{-4N}\right] .
\end{equation}

Let us set
\begin{equation}
Z ={1+\beta^2\over 1-\beta^2 }\left({2Q^2 e^{2\beta\varphi_{0}} \over \gamma^2 } - 1 \right) .
\end{equation}

The three cases  $Z>0$, $Z<0$ and $Z=0$ should be considered separately.

\subsection*{\bf Solutions with $Z>0$}

In this case no cosmological horizon is present and choosing

\begin{equation}
\lambda_{0}(r) = {r^{2-2N}\over \gamma^2} Z
\end{equation}

the mass formula (\ref{QLM}) gives

\begin{equation}
M = {1\over 2} \beta A .
\end{equation}

First we consider the case $\beta A>0$. For $0< r<\infty$ the zeros of $\lambda(r)$  are governed  by the function $\eta(r)$

\begin{equation}
\eta(r) = - {2M(1+\beta^2) \over \beta^2} + Zr + 2\left({1+ \beta^2\over 1-3\beta^2 }\right)^2 {P^2e^{-2\beta\varphi_{0} }\over \gamma^2} r^{1-4N}.
\end{equation}

The same considerations as for the function $f(r)$ show that the function $\eta(r)$ has two zeros for $\beta^2 >1/3$ when

\begin{equation}\label{MCE1}
M > {2\beta^4 \over (1+ \beta^2)(3\beta^2 -1)} Z \left[{2(1+\beta^2)\over 3\beta^2 -1 } {P^2e^{-2\beta\varphi_{0}}\over Z\gamma^2 } \right]^{1 +\beta^2\over 4\beta^2 }
\end{equation}

and  only one zero for $\beta^2<1/3$.

Therefore, the black holes with  with $\beta^2>1/3$ have two
horizons when (\ref{MCE1}) is satisfied while those with $\beta^2<1/3$ have only one horizon. The radii of the horizons can not be expressed explicitly for arbitrary
$\beta$. The extremal case corresponds to

\begin{equation}\label{MCE11}
M = {2\beta^4 \over (1+ \beta^2)(3\beta^2 -1)} Z \left[{2(1+\beta^2)\over 3\beta^2 -1 } {P^2e^{-2\beta\varphi_{0}}\over Z\gamma^2 } \right]^{1 +\beta^2\over 4\beta^2 }
\end{equation}

and is characterized with a horizon radius

\begin{equation}
r_{ext} = \left[{2(1+\beta^2)\over  3\beta^2 -1} {P^2 e^{-2\beta\varphi_{0}}\over Z\gamma^2}
\right]^{1+\beta^2\over 4\beta^2} .
\end{equation}

The solutions which do not satisfy (\ref{MCE1}) and (\ref{MCE11}) and those with $\beta A\le 0$ describe
naked singularities.

\subsection*{Solutions with $Z<0$}

For large enough values of $r$ we have $\lambda(r) \approx r^{2-2N}Z/\gamma^2 <0$ and therefore there is
a cosmological horizon in the case under consideration.

\subsection*{Solutions with $Z=0$}

First we consider the solution with $\beta^2<1/3$. In this case we have a  horizon at

\begin{equation}
r_{h} = \left[ {(1-3\beta^2)^2 \over 2\beta (1 + \beta^2) } {A\over P^2 e^{-2\beta\varphi_{0}}} \right]^{1+\beta^2 \over 1-3\beta^2 }
\end{equation}

when ${A\over \beta}>0$. For ${A\over \beta}\le 0$ we have a naked singularity.

The mass calculated via (\ref{QLM}) is $M={1\over 2}\beta A$ where the background function is given by
\begin{equation}
\lambda_{0}(r) = 2 \left({1+\beta^2\over 1-3\beta^2 } \right)^2
{P^2e^{-2\beta\varphi_{0}}\over \gamma^4}  r^{2-6N}.
\end{equation}

The black hole temperature and entropy are given by

\begin{equation}
T = {(1 - 3\beta^2) M\over 2\pi\beta^2\gamma^2 } r_{h}^{-{2\beta^2\over \ 1+\beta^2}} ,
\end{equation}

\begin{equation}
S = \pi \gamma^2 r_{h}^{2N} .
\end{equation}

For $\beta^2>1/3$ the solution has a cosmological horizon when $A>0$  while for $A\le 0$ the solution describes a naked singularity.

\section{Topological black holes}

When the asymptotic flatness is violated black holes can have nontrivial topology. Such black holes
called topological black holes attracted much interest (see for example \cite{ABHP}- \cite{B}).

The spacetime metric is given by

\begin{equation}\label{TSTM}
ds^2 = -\lambda(r)dt^2 + {dr^2\over \lambda(r)} + R^2(r){\cal {G}}_{ab}dx^adx^b
\end{equation}

where ${\cal {G}}_{ab}$ is the metric of a closed orientable  $2$-dimensional hypersurface $\Sigma_{\kappa}$ with a constant curvature $2\kappa$ $(\kappa =-1,0, 1)$ and area $\sigma_{\kappa}$. For $\kappa=1$, the spacetime metric describes black holes with two-sphere topology. For $\kappa=0$,  the topology of $\Sigma_{\kappa}$ is that of a torus. For $\kappa=-1$, the hypersurface is a two-surface of constant negative curvature and genus ${\it g}>1$ as
$\sigma_{-1} = 4\pi({\it g}-1) $.

For the metric (\ref{TSTM}) the EMd equations are reduced to

\begin{eqnarray}
{1\over R} {d^2R\over  dr^2 }  &=&
- \left({d\varphi\over dr} \right)^2 , \\
{1\over R^2} {d\over dr } \left( \lambda {dR^2\over dr}\right) &=& {2\kappa\over R^2} - {2\over R^4}\left(Q^2 e^{2\beta\varphi} + P^2 e^{-2\beta\varphi} \right) - V(\varphi) ,\\
{1\over R^2} {d\over dr }\left(R^2\lambda {d\varphi\over dr } \right) &=& {1\over 4} {dV(\varphi)\over d\varphi } +
{\beta\over R^4}\left(Q^2 e^{2\beta\varphi} - P^2 e^{-2\beta\varphi} \right).
\end{eqnarray}

where

\begin{equation}
Q = {4\pi q\over \sigma_{\kappa}} ,\,\,\, P = {4\pi p\over \sigma_{\kappa} }
\end{equation}

and $q$ and $p$ are the electric and the magnetic charge, respectively.

The black holes with $\kappa=1$ were considered in the previous sections. Here we consider the cases $\kappa=-1$
and $\kappa=0$.  It turns out that when  $\kappa=-1$, there is no dyonic solutions for the  ansatz (\ref{ansatz}).

For $\kappa=0$, the EMd equations with a Liouville potential $V(\varphi)= \Lambda e^{2k\varphi} $  have the following  dyonic solution:
\begin{eqnarray}
N &=& {1\over 1 + \beta^2} ,\\
k &=& {\beta^2 - 2\over \beta},\\
\Lambda &=& {2\beta^2 \over 1 - \beta^2 } {Q^2e^{{4\over \beta}\varphi_{0}}\over \gamma^4 }, \\
\varphi(r) &=& {\beta\over 1 + \beta^2} \ln(r) + \varphi_{0}, \\
\lambda(r) &=& {r^{-2N} \over  \gamma^2} \left[- {(1 + \beta^2)^2\over (1-\beta^2)(3\beta^2 - 1)} {Q^2 e^{2\beta\varphi_{0}} \over \gamma^2} r^{4-4N} + (1+\beta^2) {P^2 e^{-2\beta\varphi_{0}} \over \gamma^2 } + {A(1+ \beta^2)\over \beta }r\right]
\end{eqnarray}

where $A$ and $\varphi_{0}$ are constants. The solution is ill defined for $\beta^2=1/3$ and $\beta^2=1$. The discrete transformation $Q\longleftrightarrow P$, $\varphi \longleftrightarrow - \varphi$, $k \longleftrightarrow -k$ generates another solution with the same spacetime metric.

The scalar curvature is given by

\begin{equation}
{\cal R} = {2\beta^2 \over 1+ \beta^2} {r^{-2N}\over \gamma^2 } \left[ {A\over \beta r }  + {P^2e^{-2\beta\varphi_{0}}\over \gamma^2 r^2 } + 3{(1 + \beta^2)(2\beta^2 -1)\over (1-\beta^2)} {Q^2e^{2\beta\varphi_{0}} \over \gamma^2} r^{2 -4N}\right] .
\end{equation}

When  ${A\over \beta} >0 $ the solution describes a naked singularities for $\beta^2<1/3$ and $\beta^2>1$ while
for $1/3<\beta^2 <1$ there is a cosmological horizon.

In the case $A=0$ the solutions with $\beta^2<1/3$ and $\beta^2>1$ are naked singularities while for $1/3<\beta^2 <1$ there is a cosmological horizon at

\begin{equation}
r_{CH} = \left[ {(1-\beta^2)(3\beta^2-1)\over (1+\beta^2) } {P^2\over Q^2 } e^{-4\beta\varphi_{0}}\right]^{1+\beta^2\over 4\beta^2 } .
\end{equation}

As it can be seen, for ${A\over \beta }<0$ and $\beta^2<1$, there is a cosmological horizon. The most
interesting case is when ${A\over \beta }<0$ and $\beta^2>1$. The mass \footnote{In the case under consideration the spacetime topology is $R^2\times T^2$
and the quasilocal mass formula (\ref{QLM}) is modified since the integration is performed over the torus $T^2$ not over the two-sphere. This can be seen from the integral formulas given in \cite{BY}.} is found to be

\begin{equation}
m = - {\sigma_{0}\over 4\pi} {A\over 2\beta}= {\sigma_{0}\over 4\pi}M .
\end{equation}

The zeros of the metric function $\lambda(r)$
are governed by the function

\begin{equation}
\chi(r) = - {(1 + \beta^2)^2\over (1-\beta^2)(3\beta^2 - 1)} {Q^2 e^{2\beta\varphi_{0}} \over \gamma^2} r^{4-4N} + (1+\beta^2) {P^2 e^{-2\beta\varphi_{0}} \over \gamma^2 }  -2(1+ \beta^2)Mr
\end{equation}

for $0<r<\infty$.
This function satisfies $\chi(0) = +\infty$ and $\chi(\infty)=+\infty$. Besides $\chi(r)$ has only one minimum at

\begin{equation}
r_{min} = \left[ M {(\beta^2-1)(3\beta^2-1)\over 2\beta^2 {Q^2e^{2\beta\varphi_{0}}\over \gamma^2 } }\right]^{1 + \beta^2\over 3\beta^2 - 1} .
\end{equation}

The function $\chi(r)$ (respectively $\lambda(r)$) has zeros only when $\chi(r_{min})\le 0$ which gives

\begin{equation}\label{TMCE}
M\ge {2\beta^2 \over 3\beta^2 - 1} {P^2e^{-2\beta\varphi_{0}} \over \gamma^2} \left[{Q^2e^{4\beta\varphi_{0}} \over P^2(\beta^2 -1) } \right]^{1 +\beta^2\over 4\beta^2} .
\end{equation}

There are two zeros for $\chi(r_{min})<0$ and only one (degenerate) zero which corresponds to the extremal solution. Therefore, when ${A\over \beta }<0$ and $\beta^2>1$, the solution describes black holes with two horizons or one
degenerate horizon hiding a singularity at $r=0$ in both cases. The radius of the extremal solution is found explicitly $r_{ext}=r_{min}$. Let us mention that the topological dyonic  black holes exist only for negative
dilaton potential.
The solutions with ${A\over \beta }<0$ and $\beta^2>1$ which do not satisfy (\ref{TMCE}) are naked singularities.

\section{Black hole solutions in $n$-dimensions }

Here we present exact black hole solutions in $n$-dimensional\footnote{We consider $n$-dimensional spacetimes with $n>4$.} spacetimes generalizing in this way our results
for four dimensions. Following \cite{GM} we consider a dilaton gravity model described by the action

\begin{equation}\label{NDA}
{\cal S} = \int d^nx\sqrt{-g} [{\cal R}
- {4\over n-2}g^{\mu\nu}\partial_{\mu}\varphi \partial_{\nu}\varphi
- e^{-{4\beta\over n-2}\varphi}F_{\mu\nu} F^{\mu\nu}  - {2\over (n-2)! }e^{- {4\beta\over n-2}\varphi} {\cal F}_{\mu_{1}....\mu_{n-2}}  {\cal F}^{\mu_{1}....\mu_{n-2}} - V(\varphi)]
\end{equation}

where $F_{\mu\nu}$ is the field strength of the electromagnetic field and ${\cal F}_{\mu_{1}....\mu_{n-2}}$
is the $(n-2)$-form field strength of an Abelian gauge field. The metric of  $n$-dimensional spherically symmetric
spacetime can be written in the form

\begin{equation}
ds^2 = - \lambda(r)dt^2 + {dr^2\over \lambda(r) } + R^2(r)d\Omega^2_{n-2}
\end{equation}

where $d\Omega^2_{n-2}$ is the metric on the $n-2$-dimensional sphere. We consider field configurations for which the electromagnetic field has the form

\begin{equation}
F = e^{{4\beta\over n-2 }\varphi} {Q\over R^{n-2}(r) }dt\wedge dr
\end{equation}
and the Abelian gauge field is presented by

\begin{equation}
{\cal F}_{\mu_{1}...\mu_{n-2}} = {P\over R^{n-2}(r)} \varepsilon_{\mu_{1}...\mu_{n-2}}
\end{equation}

with $Q$ and $P$ being electric and magnetic charges, respectively.

The field equations derived form the action (\ref{NDA}) are

\begin{eqnarray}
{d\over dr } \left(\lambda{d\over dr }R^{n-2}\right) &=& (n-2)(n-3)R^{n-4} - {2Q^2\over R^{n-2}}e^{{4\beta\over n-2 }\varphi}-
{2P^2\over R^{n-2}}e^{-{4\beta\over n-2 }\varphi} - V(\varphi)R^{n-2}, \\
{1\over R } {d^2\over dr^2}R  &=& - {4 \over (n-2)^2 } \left({d\varphi\over dr} \right)^2 , \\
{d\over dr} \left(R^{n-2}\lambda {d\varphi\over dr } \right) &=& {n-2\over 8 } {dV(\varphi)\over d\varphi }R^{n-2}
 + \beta {Q^2 \over R^{n-2}} e^{{4\beta\over n-2 }\varphi} -  \beta {P^2 \over R^{n-2}} e^{-{4\beta\over n-2 }\varphi} .
\end{eqnarray}

We look for solutions in the form

\begin{equation}
R(r) = \gamma r^{N}
\end{equation}

where $\gamma$ and $N$ are constants.

\subsection{$n$-dimensional black holes with $V(\varphi)=0$}

We have found the following  solution

\begin{eqnarray}
\beta^2 &=& n-3, \\
N &=& {1\over n-2}, \\
\varphi_{0} &=& - {(n-2)\over 4\beta} \ln\left({4Q^2\over (n-2)(n-3)\gamma^{2(n-3)}} \right),\\
\varphi(r) &=& {\beta\over 2}\ln(r) + \varphi_{0}, \\
\lambda(r) &=& - {4M\over \gamma^{n-2} }  + {(n-2)^2\over 4\gamma^2 } r^{2{n-3\over n-2}} +
{4Q^2P^2\over (n-3)^2 \gamma^{2(2n-5)} } r^{-2{n-3\over n-2 }}.
\end{eqnarray}

Here $M$ is the asymptotic value of the quasilocal mass which. Here and below we shall consider only positive M.  Another solutions with same spacetime metric can obtained via the discrete transformation $Q\longleftrightarrow P$ , $\varphi \longleftrightarrow - \varphi$.

For

\begin{equation}
M^2 > {1\over 4} \left({n-2\over n-3 }\right)^2 {Q^2P^2\over \gamma^{2(n-2)} }
\end{equation}

there are two horizons located at

\begin{equation}
r_{\pm} = \left[{8\over (n-2)^2 \gamma^{n-4}}\left[M \pm \sqrt{M^2 - {1\over 4}\left({n-2\over n-3 }\right)^2
{Q^2P^2\over \gamma^{2(n-2)}}  } \right] \right]^{n-2\over 2(n-3) }.
\end{equation}

The extremal black holes correspond to

\begin{equation}
M = {1\over 2} {n-2\over n-3 } {|QP|\over\gamma^{n-2} }.
\end{equation}

The solutions with

\begin{equation}
M^2 < {1\over 4} \left({n-2\over n-3 }\right)^2 {Q^2P^2\over \gamma^{2(n-2)} }
\end{equation}

are naked singularities.

\subsection{$n$-dimensional black holes with a Liouville potential }

There are two classes of $n$-dimensional solutions with a Liouville potential $V(\varphi)= \Lambda e^{2k\varphi}$ where $\Lambda$ and $k$ are constants.

The first class solutions is given by

\begin{eqnarray}
\beta^2 &=& n-3 ,\\
N &=& {1\over n-2 },\\
k &=& - {2\beta\over n-2},\\
\varphi_{0} &=& - {(n-2)\over 4\beta} \ln \left({4Q^2\over (n-2)(n-3)\gamma^{2(n-3)} } \right) ,\\
\varphi(r) &=& {\beta\over 2}\ln(r) + \varphi_{0}, \\
\lambda(r) &=& - {4M\over \gamma^{n-2} }  + {(n-2)^2\over 4\gamma^2 } r^{2{n-3\over n-2}} +
{4Q^2P^2\over (n-3)^2 \gamma^{2(2n-5)} } r^{-2{n-3\over n-2 }}  - {2\Lambda Q^2 \over (n-3)\gamma^{2(n-3)}} r^{2\over n-2} .
\end{eqnarray}

The discrete transformation $Q\longleftrightarrow P$, $\varphi \longleftrightarrow -\varphi$, $k \longleftrightarrow -k$ generates another solutions with the same spacetime metric.

Unfortunately, it is not possible to solve explicitly the algebraic equation $\lambda(r)=0$ and to find the location
of the horizons. However, the causal structure can be described qualitatively. Considerations similar to those
presented in the previous sections show that there can be two non-degenerate horizons or one degenerate horizon.

The second class solutions is

\begin{eqnarray}
N &=& {\beta^2 \over (n-3)^2 + \beta^2},\\
k &=& 2\beta {(2n-5)\over (n-2)(n-3)}, \\
\Lambda &=&  - {2(n-3- \beta^2)(n-3) \over [(n-3)^2 - \beta^2(2n-5)] } \left[{(n-2)(n-3)^2\over n-3 + \beta^2 } \right]^{n-2\over n-3 } {Q^2\over (2P^2)^{n-2\over n-3 } }, \\
\varphi_{0} &=& {(n-2)\over 4\beta} \ln\left[{2P^2(n-3 + \beta^2)\over (n-2)(n-3)^2\gamma^{2(n-3)}} \right],\\
\varphi(r) &=& - {\beta\over 2} {(n-2)(n-3)\over [(n-3)^2 + \beta^2]} \ln(r) + \varphi_{0} ,\\
\lambda(r) &=& - {4M\over (n-2)N \gamma^{n-2} } r^{1 - (n-2)N} + {1\over \gamma^2 } \left[{(n-3)^2 + \beta^2\over n-3 +\beta^2 } \right]^2 r^{2(1-N)} + \nonumber \\
&+& {4Q^2P^2\over \gamma^{2(2n-5)} } {[(n-3)^2 + \beta^2]^2 (n-3 + \beta^2) \over (n-2)(n-3)^3 (n-3 - 3\beta^2) [(n-3)^2 -\beta^2(2n-5)] } r^{2[1 - (2n-5)N]} .
\end{eqnarray}

Another solutions with the same spacetime metric can be obtained via the above given discrete transformation.
The solutions are ill defined for $\beta^2= (n-3)/3$ and $\beta^2=(n-3)^2/(2n-5)$. In order to investigate the causal
structure of the solutions we introduce the function $f(r)$ defined by

\begin{equation}
\lambda(r) = r^{1-(n-2)N}f(r).
\end{equation}

For $\beta^2< {(n-3)^2\over (2n-5)}$ we have $f(r)>0$ when  $r$ is  sufficiently large and $f(r)<0$ when $r$ is sufficiently
small. In addition, ${d\over dr}f(r)>0$ and therefore $\lambda(r)$ has only one non-degenerate zero, i.e. there is one non-degenerate horizon.

For $\beta^2> {(n-3)^2\over (2n-5)}$  we can show (in the same way as for the case $n=4$)
that there exist two horizons (or one degenerate horizon) only when the following is satisfied

\begin{equation}
{M\over \gamma^{n-4}}\geq {(n-2)N\beta^2 \over 3\beta^2 -(n-3)} \left[{(n-3)^2 +\beta^2 \over n-3 +\beta^2 }\right]^2
\left[{4Q^2P^2\over \gamma^{4(n-3) }} {(n-3 + \beta^2)^2\over (n-2)(n-3)^2[\beta^2(2n-5)- (n-3)^2 ] }  \right]^{1 + (n-4)N\over 4(n-3)N } .
\end{equation}

The equality stands for the extremal solutions.

\subsection{$n$-dimensional black holes with two Liouville potentials }

Here we present a $n$-dimensional solution with $V(\varphi) = \Lambda_{1}e^{2k_{1}\varphi} + \Lambda_{2}e^{2k_{2}\varphi}$. The solution is

\begin{eqnarray}
N &=& {\beta^2 \over (n-3)^2 + \beta^2} ,\\
k_{1} &=& {2\beta\over (n-2)(n-3) }, \\
k_{2} &=& 2\beta {(2n-5)\over (n-2)(n-3) }, \\
\Lambda_{1} &=& {1\over \gamma^2} {(n-3)e^{-{4\beta\varphi_{0}\over (n-2)(n-3)}}\over [(n-3)^2 - \beta^2] }
\left[(n-2)(n-3)  - {2P^2 (n-3 + \beta^2)\over \gamma^{2(n-3)}} e^{-{4\beta\varphi_{0}\over n-2}}\right],\\
\Lambda_{2} &=& - {2Q^2 \over \gamma^{2(n-2)}} {(n-3)(n-3 - \beta^2)e^{-{4\beta\varphi_{0}\over n-3 }}\over [(n-3)^2 - \beta^2(2n-5)]},\\
\varphi(r) &=& - {\beta\over 2} {(n-2)(n-3)\over [(n-3)^2 + \beta^2]}\ln(r)+ \varphi_{0},\\
\lambda(r) &=& - {4M\over (n-2)N \gamma^{n-2}} r^{1 -(n-2)N} + {2Q^2 e^{4\beta\varphi_{0}\over n-2} \over \gamma^{2(n-2)}} {[(n-3)^2 + \beta^2]^2  r^{2[1-(2n-5)N]}\over (n-3)(n-3 - 3\beta^2) [(n-3)^2 - \beta^2(2n-5)] }
\nonumber \\ &-&
{1\over \gamma^2} {[(n-3)^2 + \beta^2]^2 \over (n-3+ \beta^2) [(n-3)^2 - \beta^2] } \left[1 - {2P^2e^{-{4\beta\varphi_{0} \over n-2}} \over \gamma^{2(n-3)} } \right] r^{2(1-N)}
\end{eqnarray}

Another solution with the same spacetime metric can be obtained by the discrete transformation
$Q\longleftrightarrow P$, $\varphi \longleftrightarrow -\varphi$, $k_{1} \longleftrightarrow - k_{1}$,
$k_{2} \longleftrightarrow - k_{2}$. The solution is ill defined for $\beta^2= (n-3)/3$, $\beta^2=(n-3)^2$ and $\beta^2= (n-3)^2/(2n-5)$.

Let us define

\begin{equation}
Z = {[(n-3)^2 +\beta^2]^2\over (n-3 + \beta^2)[(n-3)^2 - \beta^2]} \left[{2P^2 e^{-{4\beta\varphi_{0}\over n-2}} \over \gamma^{2(n-3)} } -1 \right].
\end{equation}

The three cases $Z>0$, $Z=0$ and $Z<0$ should be considered separately.

\subsection*{\bf Solutions with $Z>0$}

For $\beta^2<(n-3)^2/(2n-5)$ there is one non-degenerate horizon. For $\beta^2>(n-3)^2/(2n-5)$
there are two non-degenerate horizons or one degenerate horizon when $M$ satisfies

\begin{equation}
{M\over \gamma^{n-4}}\geq {(n-2)N\beta^2\over 3\beta^2+ 3-n }Z \left[{2Q^2 e^{4\beta\varphi_{0}\over n-2 } \over \gamma^{2(n-3)}Z } {[(n-3)^2 + \beta^2]^2\over (n-3 +\beta^2) [\beta^2(2n-5) - (n-3)^2]} \right]^{1 +(n-4)N\over 4(n-3)N}
\end{equation}

as the equality corresponds to the extremal case.

\subsection*{\bf Solutions with $Z=0$}

For $\beta^2<(n-3)/3$ there is one non-degenerate horizon with

\begin{equation}
r_{h} = \left[{(n-2)[(n-3)^2 + \beta^2]^2 N  \over 2(n-3)(n-3 - 3\beta^2)[(n-3)^2 - \beta^2(2n-5)] } {Q^2e^{4\beta\varphi_{0}\over n-2}   \over M\gamma^{n-2} }\right]^{(n-3)^2+\beta^2 \over (n-3)(3\beta^2 -n+3) }.
\end{equation}

For $\beta^2>(n-3)/3$ the solution has a cosmological horizon.

\subsection*{\bf Solutions with $Z<0$}

The solutions with $Z<0$ have a cosmological horizon.

\section{Conclusion}

In summary, we have obtained  several families of spherically symmetric dyonic black hole solutions in  four dimensional and higher dimensional EMd gravity without a dilaton  potential or with a Liouville type  potentials. These black holes have unusual asymptotics. They are neither asymptotically flat nor asymptotically (anti-) de Sitter. A class of non-asymptotically flat, non-dS/AdS topological dyonic black holes with a toroidal horizon has also been presented in four dimensions. Some basic properties of the black holes have been discussed.

The  black hole solutions presented here break all supersymmetries and therefore, they, as examples of non-asymptotically flat and non-AdS/dS solutions, may serve as  backgrounds   for  non-supersymmetric holography and they might lead to possible extensions of AdS/CFT correspondence as it is stated in  \cite{CGL}, \cite{CL},
\cite{CMZ}, \cite{CO} and\cite{CZ1} .

\section*{Acknowledgements} I would like to thank V. Rizov and B. Slavov for reading the manuscript.
This work was partially supported by the Bulgarian National Research Fund.

\end{document}